\documentclass{egpubl}
\usepackage{eurovis2025}

\SpecialIssueSubmission    % uncomment for submission to , special issue
\CGFccby
\usepackage[T1]{fontenc}
\usepackage{dfadobe}  

\usepackage{cite}  % comment out for biblatex with backend=biber
\BibtexOrBiblatex
\electronicVersion
\PrintedOrElectronic
\ifpdf \usepackage[pdftex]{graphicx} \pdfcompresslevel=9
\else \usepackage[dvips]{graphicx} \fi

\usepackage{egweblnk}
\title[Representing Visualization Insights as a Dense Insight Network]%
      {Representing Visualization Insights as a Dense Insight Network}

\author[Hoffswell \etal]
{\parbox{\textwidth}{\centering% 
    Jane Hoffswell$^{1}$,
    Victor Soares Bursztyn$^{1}$,
    Shunan Guo$^{1}$,
    Jesse Martinez$^{2}$,
    and Eunyee Koh$^{1}$
        }
        \\
{\parbox{\textwidth}{\centering%
$^1$Adobe Research, USA\\ 
$^2$University of Washington, USA
       }
}
}
\usepackage{xargs}      % Used for new commands with optional arguments
\usepackage{soul}       % Used for custom comments
\usepackage{color}      % Used for custom colors in comments
\usepackage{xspace}     % Used for abbreviation spacing
\usepackage{xpunctuate} % Used for abbreviation spacing
\usepackage{tcolorbox}  % Used for the colorbox labels
\usepackage{xurl}       % Used to support url linebreaks in the footnote
\usepackage{mdframed}   % Used to display our LLM prompt
\usepackage{array}      % Used to shrink the minimum column size

\newcommand{\ie}{{i.e.,}\xspace}
\newcommand{\eg}{{e.g.,}\xspace}
\newcommand{\etal}{{et~al\xperiod}\xspace}

\newcommand{\etc}{{etc\xperiod}\xspace}

\renewenvironment{quote}{%
  \list{}{%
    \leftmargin0.5cm   % this is the adjusting screw
    \rightmargin\leftmargin
  }
  \item\relax
}
{\endlist}

\newcommand{\myquote}[1]{\emph{``#1''}}                         % Italic text in quotation marks
\newcommand{\insight}[1]{\emph{``#1''}}

\newcommand{\bpstart}[1]{\vspace{4px}\noindent{\textbf{#1.}}}

\definecolor{lightpink}{RGB}{237,157,202}
\definecolor{lightred}{RGB}{210,121,121}
\definecolor{lightorange}{RGB}{230,170,50}
\definecolor{lightgold}{RGB}{210,194,121}
\definecolor{lightgreen}{RGB}{121,210,121}
\definecolor{lightaqua}{RGB}{121,206,210}
\definecolor{lightblue}{RGB}{121,124,210}
\definecolor{NavyBlue}{RGB}{0,0,0}
\definecolor{lightpurple}{RGB}{153,102,255}
\definecolor{red}{RGB}{178,34,34}
\definecolor{gray}{RGB}{166,166,166}

\definecolor{dashRed}{RGB}{236, 63, 47}
\definecolor{dashOra}{RGB}{240,107, 24}
\definecolor{dashYel}{RGB}{231,186, 82}
\definecolor{dashGre}{RGB}{ 70,171, 94}
\definecolor{dashTea}{RGB}{  0,128,128}
\definecolor{dashBlu}{RGB}{ 69,147,196}
\definecolor{dashPur}{RGB}{108, 88,166}
\definecolor{dashPin}{RGB}{236, 83,157}

\newtcbox{\dashWhi}{nobeforeafter,tcbox raise base,                       % inline
boxrule=0mm,top=0mm,bottom=0mm,right=0mm,left=0mm,arc=0.5mm,boxsep=0.6mm, % padding/box
fontupper=\sffamily\small\bfseries,                                       % font
colframe=white,coltext=black,colback=white}                           % color

\newtcbox{\dashGra}{nobeforeafter,tcbox raise base,                       % inline
boxrule=0mm,top=0mm,bottom=0mm,right=0mm,left=0mm,arc=0.5mm,boxsep=0.6mm, % padding/box
fontupper=\sffamily\small\bfseries,                                       % font
colframe=gray,coltext=white,colback=gray}                           % color

\newtcbox{\dashRed}{nobeforeafter,tcbox raise base,                       % inline
boxrule=0mm,top=0mm,bottom=0mm,right=0mm,left=0mm,arc=0.5mm,boxsep=0.6mm, % padding/box
fontupper=\sffamily\small\bfseries,                                       % font
colframe=dashRed,coltext=white,colback=dashRed}                           % color

\newtcbox{\dashOra}{nobeforeafter,tcbox raise base,                       % inline
boxrule=0mm,top=0mm,bottom=0mm,right=0mm,left=0mm,arc=0.5mm,boxsep=0.6mm, % padding/box
fontupper=\sffamily\small\bfseries,                                       % font
colframe=dashOra,coltext=white,colback=dashOra}                           % color

\newtcbox{\dashYel}{nobeforeafter,tcbox raise base,                       % inline
boxrule=0mm,top=0mm,bottom=0mm,right=0mm,left=0mm,arc=0.5mm,boxsep=0.6mm, % padding/box
fontupper=\sffamily\small\bfseries,                                       % font
colframe=dashYel,coltext=white,colback=dashYel}                           % color

\newtcbox{\dashGre}{nobeforeafter,tcbox raise base,                       % inline
boxrule=0mm,top=0mm,bottom=0mm,right=0mm,left=0mm,arc=0.5mm,boxsep=0.6mm, % padding/box
fontupper=\sffamily\small\bfseries,                                       % font
colframe=dashGre,coltext=white,colback=dashGre}                           % color

\newtcbox{\dashTea}{nobeforeafter,tcbox raise base,                       % inline
boxrule=0mm,top=0mm,bottom=0mm,right=0mm,left=0mm,arc=0.5mm,boxsep=0.6mm, % padding/box
fontupper=\sffamily\small\bfseries,                                       % font
colframe=dashTea,coltext=white,colback=dashTea}                           % color

\newtcbox{\dashBlu}{nobeforeafter,tcbox raise base,                       % inline
boxrule=0mm,top=0mm,bottom=0mm,right=0mm,left=0mm,arc=0.5mm,boxsep=0.6mm, % padding/box
fontupper=\sffamily\small\bfseries,                                       % font
colframe=dashBlu,coltext=white,colback=dashBlu}                           % color

\newtcbox{\dashBluOut}{nobeforeafter,tcbox raise base,                       % inline
boxrule=0.3mm,top=0mm,bottom=0mm,right=0mm,left=0mm,arc=0.5mm,boxsep=0.2mm, % padding/box
fontupper=\sffamily\small\bfseries,                                       % font
colframe=dashBlu,coltext=black,colback=white}                           % color

\newtcbox{\dashPur}{nobeforeafter,tcbox raise base,                       % inline
boxrule=0mm,top=0mm,bottom=0mm,right=0mm,left=0mm,arc=0.5mm,boxsep=0.6mm, % padding/box
fontupper=\sffamily\small\bfseries,                                       % font
colframe=dashPur,coltext=white,colback=dashPur}                           % color

\newtcbox{\dashPin}{nobeforeafter,tcbox raise base,                       % inline
boxrule=0mm,top=0mm,bottom=0mm,right=0mm,left=0mm,arc=0.5mm,boxsep=0.6mm, % padding/box
fontupper=\sffamily\small\bfseries,                                       % font
colframe=dashPin,coltext=white,colback=dashPin}                           % color

\begin{document}

% uncomment for using teaser
\teaser{
 \includegraphics[width=0.98\linewidth]{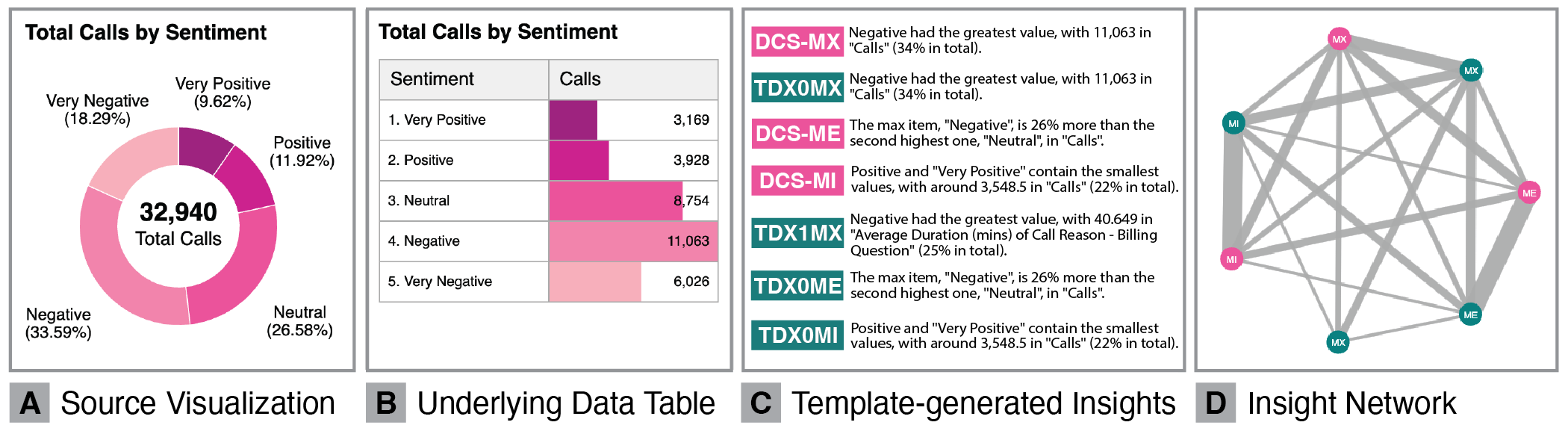}
 \centering
  \caption{An overview of the insight generation pipeline for our insight network framework. (A) Starting with a visualization from the~source dashboard, (B) we represent the data internally as a table and (C) generate a set of template-based insights (\dashPin{DCS-MX}\,,\,\dashPin{DCS-ME}\,,\,\dashPin{DCS-MI}). Along with the three insights generated for the donut chart, we also show four related insights (\dashTea{TDX0MX}\,,\,\dashTea{TDX1MX}\,,\,\dashTea{TDX0ME}\,,\,\dashTea{TDX0MI}) generated based on the table in sub-panel \dashTea{D} in the original dashboard (Figure~\ref{fig:dashboard}). (D) We then encode the characteristics that connect these insights using our insight network framework; the network shown here is a simplification for this subset of seven insights.}
\label{fig:teaser}
}

\maketitle
%-------------------------------------------------------------------------
\begin{abstract}
   We propose a dense insight network framework to encode the relationships between automatically generated insights from a complex dashboard based on their shared characteristics. Our insight network framework includes five high-level categories of relationships (e.g.,~type, topic, value, metadata, and compound scores). The goal of this insight network framework is to provide a foundation for implementing new insight interpretation and exploration strategies, including both user-driven and automated approaches. To illustrate the complexity and flexibility of our framework, we first describe a visualization playground to directly visualize key network characteristics; this playground also demonstrates potential interactive capabilities for decomposing the dense insight network. Then, we discuss a case study application for ranking insights based on the underlying network characteristics captured by our framework, before prompting a large language model to generate a concise, natural language~summary. Finally, we reflect on next steps for leveraging our insight network framework to design and evaluate~new~systems. 
%-------------------------------------------------------------------------
%  ACM CCS 1998
%  (see https://www.acm.org/publications/computing-classification-system/1998)
% \begin{classification} % according to https://www.acm.org/publications/computing-classification-system/1998
% \CCScat{Computer Graphics}{I.3.3}{Picture/Image Generation}{Line and curve generation}
% \end{classification}
%-------------------------------------------------------------------------
%  ACM CCS 2012 (see https://www.acm.org/publications/class-2012)
%The tool at \url{http://dl.acm.org/ccs.cfm} can be used to generate
% CCS codes.
%Example:

% \ccsdesc[500]{Information systems~Data analytics}

\end{abstract}  
%-------------------------------------------------------------------------

%%%% GENERAL

\newcommand{\numInsights}{\callsGeneratedInsightsMedium}
\newcommand{\numInsightsA}{forty-nine}

%%%% TYPE-BASED

\newcommand{\insightSPa}{%
    \insight{`Calls' significantly increased in the span between Oct. 4th and 6th, growing by 10\% from 1,049 to 1,152} (from sub-panel~\dashYel{A})}
\newcommand{\insightSPb}{%
    \insight{`Very Positive' increased notably during the period of Oct. 2nd to 6th, up by 37\% from 86 to 118.} (from sub-panel~\dashBlu{E})}
\newcommand{\insightSPd}{%
    \dashTea{TDX0MI} from sub-panel~\dashTea{D} -- \insight{`Positive' and `Very Positive' contain the smallest values, with around 3,548.5 in `Calls' (22\% in total)} --}

%%%% TOPIC-BASED

\newcommand{\metricFirst}{\texttt{Calls}}
\newcommand{\metricFirstCount}{all~five}
\newcommand{\metricFirstPanels}{\dashYel{A}~\dashOra{B}~\dashPin{C}~\dashTea{D}~\dashBlu{E}}

\newcommand{\metricSecond}{\texttt{Sentiment}}
\newcommand{\metricSecondCount}{three}
\newcommand{\metricSecondPanels}{\dashPin{C}~\dashTea{D}~\dashBlu{E}}

\newcommand{\exampleTableFilters}{In our example, sub-panel~\dashTea{D} uses filters on several columns to subdivide the \texttt{Call\,Reason} into the individual categories to show the average \texttt{Duration} for each.}

\newcommand{\valueMetricNameA}{\texttt{Sentiment}}
\newcommand{\valueNameA}{``Negative''}
\newcommand{\valueCountA}{sixteen}
\newcommand{\valueMetricNameB}{\texttt{Sentiment}}
\newcommand{\valueNameB}{``Neutral''}
\newcommand{\valuePercentA}{32.7\%}
\newcommand{\valueCountB}{fourteen}
\newcommand{\valuePercentB}{28.6\%}

\newcommand{\exampleMultipleValues}{\todo{}}

%%%% TIME-BASED

\newcommand{\exampleTimePanel}{\dashBlu{E}}
\newcommand{\exampleTimeA}{\insight{`Sentiment [Negative]' grew significantly between Oct. 7th and 10th, up by 13\% from 355 to 400}}
\newcommand{\exampleTimeB}{\insight{The lowest amount of `Sentiment [Neutral]' of 255 appeared on Oct. 8th, 13\% less than the average of 291.}}

%%%% SCORE-BASED
\newcommand{\numberScores}{three}

%%%% GATEKEEPING

\newcommand{\numberTopicMetricNodes}{fifteen}
\newcommand{\numberTopicDimensionNodes}{five}
\newcommand{\panelsSegmentBreakdown}{\dashTea{D} and \dashBlu{E}}
\newcommand{\numberTopicSegmentNodes}{two}
\newcommand{\numberTopicDimensionNodesAll}{seven}
\newcommand{\numberTopicCombinedNodes}{three}
\newcommand{\exampleCommonMetricCount}{nine}
\newcommand{\exampleCommonMetric}{\texttt{Calls}}

\newcommand{\gatekeepingScores}{\texttt{priority}~(10), \texttt{layoutScore}~(9), and \texttt{valueScore}~(10)}

\newcommand{\numberInsightTypeNodes}{nine}
\newcommand{\numberTopicTotalNodes}{twenty-five}
\newcommand{\numberScoreNodes}{thirty-one}

\newcommand{\totalNodes}{116}
\newcommand{\totalLinks}{3693}

%%%% NETWORK VISUALIZATION

\newcommand{\combinedNodeOrder}{first}
\newcommand{\combinedNodeTopic}{\texttt{Sentiment}}
\newcommand{\combinedNodeInsightsA}{twenty-five}
\newcommand{\combinedNodeTopicA}{sub-panel~\dashBlu{E}}
\newcommand{\combinedNodeInsightsB}{thirteen}
\newcommand{\combinedNodeTopicB}{sub-panel~\dashTea{D}}
\newcommand{\combinedNodeInsightsC}{three}
\newcommand{\combinedNodeTopicC}{sub-panel~\dashPin{C}}
\newcommand{\combinedNodeTotal}{41}
\newcommand{\combinedNodeOtherCount}{two}
\newcommand{\combinedNodeOtherNames}{\texttt{Calls} and \texttt{Date}}
\newcommand{\combinedNodeOtherCounts}{39 and 25}

%%%% STORY EXPLORATION

\newcommand{\selectedNode}{\texttt{Calls:\,Decline} \dashYel{DE}}
\newcommand{\selectedInsight}{\insight{`Calls' significantly decreased in the span between Oct. 21st and 26th, declining by 10\% from 1,170 to 1,054,}}
\newcommand{\selectedVis}{line chart}
\newcommand{\selectedVisHighlight}{for ``Oct. 21st'' and ``Oct 26th''}

\section{Introduction}
Dashboards use visualizations and tables to provide a comprehensive overview of multiple, complex datasets~\cite{sarikaya-dashboards}. To improve interpretability by users with varying expertise, dashboards can incorporate automated captions to highlight important information in natural language~\cite{srinivasan-voder}. These insight captions are often generated individually for each visualization, which makes it difficult to browse the overarching themes and develop a clear picture of the most important takeaways using natural language alone. To complicate matters, while dashboards support non-linear exploration of the data by viewing multiple charts side-by-side, reading through the natural language insights necessitates a linear progression that can be hard to browse; furthermore, a single dashboard may give rise to many potential narrative orders depending on the goals of the viewer. In fact, effective communication has been identified as one of the major challenges faced by dashboards today~\cite{sarikaya-dashboards}.

To better support exploration of natural language insight collections, we begin this work with the research question: \emph{how are individual insights connected?} Insights have many inherent characteristics that may inform their relative importance or relationship to other insights for the same dashboard, such as the insight type or underlying data attributes. The primary goal for this work is to understand these shared characteristics in order to inform the future design and evaluation of insight ranking or selection approaches. 

We thus contribute a dense insight network framework formed of five high-level categories of connections (the links) between pairs of insights (the nodes):
(1)~\emph{type-based}~(\eg~statistical insights of the same type should be connected, such as the insights describing the max value or the ones highlighting an upward trend); 
(2)~\emph{topic-based}~(\eg~insights for a particular dimension, metric, or filter segment should be grouped); 
(3)~\emph{value-based}~(\eg~insights referencing the same dates or values should be explored together, perhaps in an ordered fashion); 
(4)~\emph{metadata-based}~(\eg~the original layout of the dashboard or other metadata should influence the reading order of insights);
and (5)~\emph{score-based}~(\eg~compound or computed scores based on the aforementioned properties can provide more complex recommendations for ordered insight~traversal).

The result of our insight network framework is an exceedingly dense ``hairball'' that is difficult to directly interpret in-and-of-itself. However, we see this complexity as an advantage, not a limitation, of our framework because it provides the flexibility to encode many different types of relationships amongst the generated insights.

To illustrate the complexity of our dense insight network, we explore several interactive network representations in a visualization playground to reflect on connection patterns and better communicate the scope of our framework. Our visualization playground displays selected insights in a threaded narrative and surfaces or highlights new insights in the interactive visualizations based on current exploration patterns. This application was designed to directly visualize the network and connections between insights, while also exploring interactive capabilities to simplify the inherent complexity in our framework. The playground is not intended as a standalone application, but rather as an exploratory communication medium.

The purpose of our dense insight network framework is to provide a foundation for building new applications or ranking strategies for insight selection. To illustrate the utility of this approach, our second research question thus asks \emph{how can we leverage our insight network framework to support automated summarization of complex dashboards?} Motivated by the inherent difficulty around reading large collections of natural language insights from a source dashboard, this case study application aims to provide a representative overview of the key information in a short summary paragraph. Informed by weekly discussions with a group of domain expert collaborators, we encode the key characteristics of a source dashboard in our insight network framework, and develop custom score-based links to rank and select a subset of insights. We then leverage a large language model~(\mbox{GPT-3.5}) to rephrase the selected insights into a more natural, concise format for the final dashboard summary.  

In summary, our primary contribution is the design of our dense \textbf{insight network framework}, which includes five high-level categories of insight connections, \ie \emph{type-}, \emph{topic-}, \emph{value-}, \emph{metadata-}, and \emph{score}-based links~(Section~\ref{sec:network}). We illustrate the complexity and flexibility of the insight network framework through a sample visualization playground that supports interactive browsing and selection of insights~(Section~\ref{sec:system}). Finally, to demonstrate the utility of our insight network framework, we develop an example \textbf{case study application} for automatic LLM-based summarization to simplify interpretation of the natural language insights from a large visualization dashboard~(Section~\ref{sec:llm}). This case study illustrates how our insight network framework can act as the foundation for building new tools and facilitating discussion of important characteristics of the dashboard and automatically-generated insights.
\begin{figure*}[t]
    \centering
    \includegraphics[width=\textwidth]{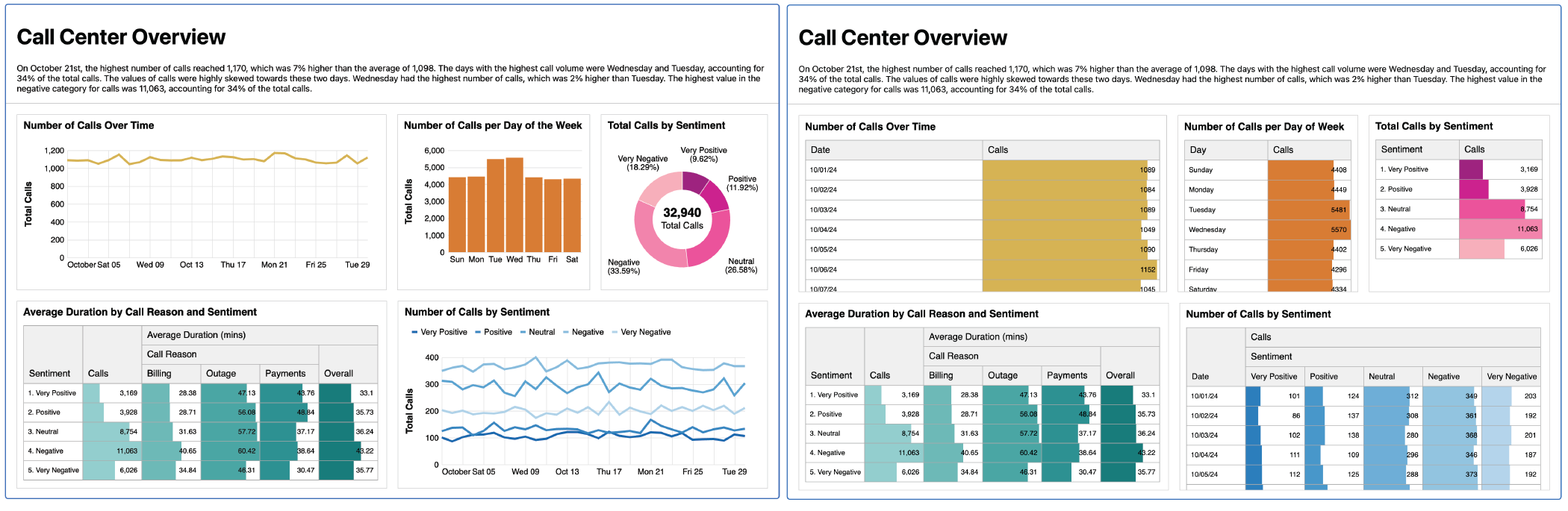}
    \vspace{-16px}
    \caption{An example dashboard titled ``Call Center Overview'' (left) and the same dashboard with the underlying data tables shown for all sub-panels instead of the visualizations (right). The dashboard has five sub-panels showing the average duration and number of calls to a call center, broken down by the sentiment and reason for the call: \dashYel{A} a line chart showing the total number of \texttt{Calls} per \texttt{Date}; \dashOra{B}~a~bar chart of the total number of \texttt{Calls} for each \texttt{Day} of the week; \dashPin{C} a donut chart of the number of \texttt{Calls} by \texttt{Sentiment}; \dashTea{D}~a~data table~showing the number of calls by \texttt{Sentiment}, as well as the average call \texttt{Duration} broken down by both call \texttt{Reason} and \texttt{Sentiment}; and \dashBlu{E}~a~multi-line chart showing the number of \texttt{Calls} by \texttt{Sentiment} per \texttt{Date}. For demonstration purposes, we generate forty-nine insights for this sample dashboard~(Section~\ref{sec:overview}). Note that the dashboard colors were chosen for illustrative purposes to match the network figures in this paper. The dashboard description shows the resulting LLM-based summary produced via our case study application~(Section~\ref{sec:llm}).}
    \label{fig:dashboard}
\end{figure*}
\section{Related Work}
This work is motivated by prior research on automatic insight generation, data storytelling and narrative artifacts, network visualization, and LLM-based insight summarization pipelines.

%%%%%%%%%%%%%%%%%%%%%%%%%%%%%%%%%%%%%%%%%%%%%%%%%%%%%%%%%%%%%%%%%%%%%%%%%%%%%
\subsection{Automated Insights and Storytelling}

Prior research has extensively explored the automatic generation of natural language insights and visualization narratives from large datasets~\cite{tang2023vistext,kantharaj2022chart,ding2019quickinsights,tang2017extracting,srinivasan-voder}. \mbox{To reduce} manual exploration and expedite data investigation, automated insight mining analyzes multidimensional data to suggest interesting patterns (i.e., data facts) that are statistically significant~\cite{ding2019quickinsights, tang2017extracting}. Battle and Ottley analyzed the common definitions for what constitutes an ``insight'' and proposed a unified formalism for describing relevant characteristics~\cite{battleOttley2023insight}; within this framing, our work focuses on the subset of insights related to \emph{data~facts}.

To facilitate exploration and comprehension, Srinivasan~\etal developed Voder~\cite{srinivasan-voder}, an interactive system that associates data facts with embellished visualizations. More recently, there has been a growing body of work exploring techniques to better organize data facts into more sophisticated narrative artifacts, such as data-driven fact sheets~\cite{wang2019datashot}, \mbox{visual} data stories~\cite{shi-calliope}, and videos~\cite{shi2021autoclips} that can facilitate both sense-making and decision-making using natural language insights. These systems typically follow a linear workflow~\cite{lee2015more} that involves uncovering insights from raw datasets, extracting or sequencing data facts that are logically connected by a coherent theme, and presenting the insights through visualizations and accompanying narratives. 

Despite the convenience and minimal effort required to generate these artifacts, these tools often restrict users' ability to influence the output by incorporating their intentions or feedback. Recent research efforts strike a balance between user control and automation by enabling user to select and arrange the automated insights in the output~\cite{zhang2022codas}, specify several key insights~\cite{sun-erato}, or engage with the the system through a question-answering interface to articulate preferences via abstract levels of~control~\cite{wu-socrates}.

While much of the aforementioned research concentrates on the end-to-end workflow from raw data to narrative visualizations, there has also been a parallel effort aimed at transforming intermediate narrative artifacts into more refined formats to improve readability. For example, ChartStory~\cite{zhao2021chartstory} converts charts into a data comic by optimizing chart grouping and layout, and generating text descriptions. NB2Slides~\cite{zheng2022telling} and Notable~\cite{li2023notable} generate presentation slides based on analysis threads in a computational notebook. The distinguishing factor is the presence of embedded analytical intentions in these intermediate narrative artifacts, which must be extracted and integrated into the final artifact. 

In this work, we focus on insight exploration and storytelling for complex dashboards, specifically on the process to automatically generate concise narrative summaries from an analytical dashboard. Despite their widespread use as analytical artifacts, the challenge of storytelling with dashboards has been underscored in prior studies~\cite{zhang-analysts, sarikaya-dashboards}, with automated generation of narrative artifacts for dashboards remaining relatively understudied. Furthermore, our proposed method draws inspiration from the need for controllability in existing automated data storytelling tools, by supporting user-guided narrative curation from diverse perspectives.

%%%%%%%%%%%%%%%%%%%%%%%%%%%%%%%%%%%%%%%%%%%%%%%%%%%%%%%%%%%%%%%%%%%%%%%%%%%%%
\subsection{Insights and Network Visualization}

While the concept of networked insights has been proposed in some prior work~\cite{smuc2009score,gotz2006interactive,willett2011commentspace,he2020characterizing,chen2009toward}, our work aims to provide a comprehensive framework for flexibly representing or composing complex relationships. Smuc~\etal~propose the idea of a relational insight organizer (RIO), which uses a row-based layout to organize insights across categories and arrows to indicate relational links when insights build upon one another~\cite{smuc2009score}. In a similar vein, several systems support manually linking insights via direct user interactions~\cite{gotz2006interactive,willett2011commentspace,he2020characterizing}. Chen~\etal analyze the categorization of data facts by insight type, and also highlight some relevant metadata that can be applied to network construction~\cite{chen2009toward}; our insight network framework expands on this idea to flesh out the connections to include additional layout-based metadata and other value-based links.

\newcommand{\exampleInsight}[1]{\footnotesize#1}
\newcommand{\insightValue}[1]{\emph{#1}}

\begin{table*}
    \centering
    \begin{tabular}{ccccccl}

    \hline
    \rule{0pt}{2.25ex}\textbf{Type} 
        & \hspace{-5px}\dashYel{A}\hspace{-5px}
        & \hspace{-5px}\dashOra{B}\hspace{-5px}
        & \hspace{-5px}\dashPin{C}\hspace{-5px}
        & \hspace{-5px}\dashTea{D}\hspace{-5px}
        & \hspace{-5px}\dashBlu{E}\hspace{-5px}
        & \textbf{Example}\\
    
    \hline
    \rule{0pt}{2.25ex}\textsc{mi} & 1 & 1 & 1 & 3 & 5 & \dashTea{TDX0MI}*
        \exampleInsight{\insightValue{Positive} and \insightValue{Very Positive} contain the smallest values, with around 3,548.5 in ``Calls'' (22\% in total).}\\
    \textsc{mx} & 1 & - & 1 & 3 & 6 & \dashYel{LCD-MX}
        \exampleInsight{``Calls'' peaked at 1,170 on Oct. 21st. This was 7\% more than the average of 1,098.}\\
    \textsc{me} & - & 1 & 1 & 5 & - & \dashPin{DCS-ME}*
        \exampleInsight{The max item, \insightValue{Negative}, is 26\% more than the second highest one, \insightValue{Neutral}, in ``Calls.''}\\
    \textsc{hb} & - & 1 & - & 2 & - & \dashTea{TDX4HB}
        \exampleInsight{\insightValue{Negative}, [and four others] contain the greatest values, with around 36.812 in ``Overall'' (100\% in total).}\\
    \textsc{sk} & - & 1 & - & - & - & \dashOra{BCW-SK}
        \exampleInsight{The values of ``Calls'' are highly skewed towards \insightValue{Tuesday} and \insightValue{Wednesday} (34\% in total).}\\
    \textsc{sp} & 1 & - & - & - & 6 & \dashYel{LCD-SP}
        \exampleInsight{``Calls'' significantly increased in the span between Oct. 4th and 6th, growing by 10\% from 1,049 to 1,152.}\\
    \textsc{de} & 1 & - & - & - & 6 & \dashBlu{MCS1DE}
        \exampleInsight{\insightValue{Positive} significantly decreased in the span between Oct. 21st and 24th, declining by 29\% from 167 to 119.}\\
    \textsc{an} & - & - & - & - & 1 & \dashBlu{MCS-AN}
        \exampleInsight{\insightValue{Positive} has the most notable anomaly [...] the value showed a 28\% difference compared to expected value.}\\
    \rule{0pt}{2.25ex}\textsc{co} & - & - & - & - & 1 & \dashBlu{MCS-CO}
        \exampleInsight{No strong correlation was identified in any series pairs.}\\
    \textsc{lt} & - & - & - & - & - & \dashGra{\,\,SALES\,\,}
        \exampleInsight{There are 25 items [...] less than or equal to 54,285.58 [...] which combined represent 28\% in total.}\\
    \textsc{se} & - & - & - & - & - & \dashGra{\,\,SALES\,\,}
        \exampleInsight{There was a cyclic pattern every 7 day(s). For each cycle, the peak [...] occurred on average on day 3 [...].}\\
    \textsc{tr} & - & - & - & - & - & \dashGra{\,\,SALES\,\,}
        \exampleInsight{The period from Nov. 16th to Dec. 24th saw an upward trend [...] going from 8,132.05 to 11,851.79 overall.}\\

    \hline
    \rule{0pt}{2.25ex}\textbf{Total} & 4 & 4 & 3 & \hspace{-5px}13\hspace{-5px} & \hspace{-5px}25\hspace{-5px} & \\
    \hline

    \end{tabular}
    \caption{We generate \numInsightsA\ insights for our sample dashboard (Figure~\ref{fig:dashboard}). This table shows the number of insights for each sub-panel and insight type, along with an example insight. Due to the particular characteristics in the data, no long tail distribution~(\textsc{lt}), seasonality~(\textsc{se}), or trend~(\textsc{tr}) insights were generated for this dashboard; the examples were thus taken from a different dashboard (``Regional Sales Summary''), which is included in the supplemental material. Several insights were edited in the interest of space; the full example insights are also included in the supplemental material. *Note: these two example insights were generated for \emph{both} the donut chart \dashPin{C} and table \dashTea{D}.}
    \label{tab:insights}
    \vspace{-6px}
\end{table*}

%%%%%%%%%%%%%%%%%%%%%%%%%%%%%%%%%%%%%%%%%%%%%%%%%%%%%%%%%%%%%%%%%%%%%%%%%%%%%
\subsection{Insight Summarization with LLMs}

The recent rise in the popularity of large language models (LLMs) has prompted different scientific communities to evaluate their potential for summarizing domain-specific content~\cite{10.1162/tacl_a_00632, goyal2022news, tang2023evaluating}. For example, InsightPilot~\cite{ma2023insightpilot} uses LLMs to generate a sequence of analysis actions by iteratively selecting the most relevant insights based on users' queries and determining subsequent analysis actions; InsightPilot ultimately produces a report with the LLM summarizing the key insights along the analysis trajectory. DataTale~\cite{sultanum2023datatales} utilizes LLMs to craft data-driven articles by integrating raw data into the prompt and linking the generated narratives back to the visualizations via keyword matching. InsightLens~\cite{weng2024insightlens} presents a multi-agent framework that organizes the data insights based on the conversational history between the user and the LLM-powered data analysis assistant. 

While LLMs show promise in generating natural-sounding summaries that appeal to human judges~\cite{10.1162/tacl_a_00632, goyal2022news, tang2023evaluating}, they are also more likely to miss or hallucinate important details such as names, dates, and percentages that convey key information, especially in fact-intensive domains~\cite{goyal2022news, 10.1145/3571730, pagnoni-etal-2021-understanding, maynez-etal-2020-faithfulness, kryscinski-etal-2020-evaluating, falke-etal-2019-ranking}. In the medical domain~\cite{tang2023evaluating}, for example, ``omission of important information'' was the number one reason reported by human judges for choosing their \emph{least} preferred LLM-generated summaries. In the data analysis domain, empirical studies have identified transparency issues in LLM-powered tasks, where users report a need for greater control over the level of assistance from the LLM and the context being provided to the LLM~\cite{gu2024data, chopra2023conversational}. Thus, striking a balance between generating natural-sounding and factually-complete summaries while providing users certain levels of control is one of the key challenges in LLM-based summarization, and hence of particular interest in our work. By leveraging a network of insights for insight selection, our approach aims to reduce the unpredictability of the LLM-generated summary by providing more guidance as to what information is most important to include via their relationships.
\section{Insight Network Framework}
\label{sec:network}

Our main contribution in this work is the design of our insight network framework, which provides a structured approach for describing the relationships between insights generated for the same dashboard. Given a set of insights (\ie~nodes), we identified five categories of insight connections, \ie links: (1)~\emph{type-based}, (2)~\emph{topic-based}, (3)~\emph{value-based}, (4)~\emph{metadata-based}, and (5)~\emph{score-based}.

In this section, we first provide some background on the process for generating our sample insights (Section~\ref{sec:overview}), though we~note that this framework could generalize to other types of insights and insight-generation approaches. Then, for each high-level link category, we introduce several of the underlying characteristics to consider, and reflect on the utility of these links for analysis and narrative exploration of the insights (Sections~\ref{sec:type}-\ref{sec:score}). 

Throughout this paper, we use the sample dashboard shown in Figure~\ref{fig:dashboard} as a consistent running example, which was creating using a ``Real World Fake Data'' dataset~\cite{callcenteroverview}. This dashboard produces \numInsightsA\ insights (\ie nodes); Table~\ref{tab:insights} shows several example insights. Figure~\ref{fig:connections} illustrates some of the node clusters (\ie cliques) in the network based on a subset of the links in our framework; we include the full link matrix in the supplemental material, which visualizes \emph{all} of the links described in this paper, with the heatmap colors showing the number of links between each pair of nodes. These visualizations are discussed in more detail in Section~\ref{sec:system}.

\newcommand{\layoutPR}{A}
\newcommand{\layoutPC}{B}
\newcommand{\layoutTC}{C}
\newcommand{\layoutPRC}{D}
\newcommand{\layoutC}{E}
\newcommand{\topicM}{F}
\newcommand{\topicD}{G}
\newcommand{\insightT}{H}
\newcommand{\scoreC}{I}
\begin{figure*}[t]
    \centering
    \includegraphics[width=\textwidth]{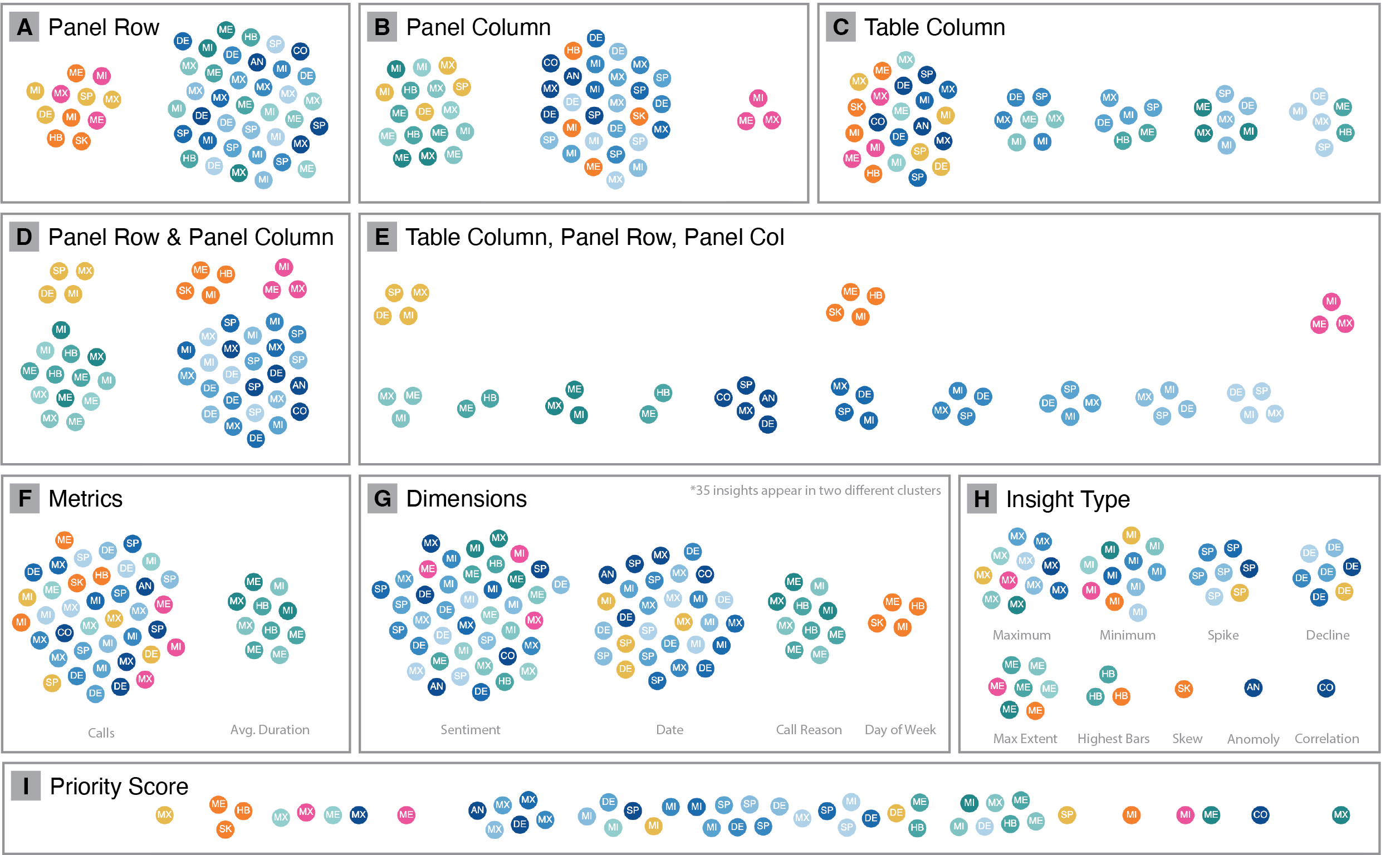}
    \vspace{-16px}
    \caption{Sample insight clusters produced in our visualization playground based on different types of links in our insight network framework. Each cluster is a clique, with links connecting every pair of nodes in the cluster. Each sub-figure only includes one node per insight unless otherwise indicated; the clustering for ``Dimensions'' (\topicD) is the notable exception, as some insights correspond to multiple dimensions.}
    \label{fig:connections}
\end{figure*}

\subsection{Background}
\label{sec:overview}

Our insight network framework encodes information about a~source dashboard, the automatically generated insights, and the associated metadata. We assume that the dashboard contains multiple sub-panels (e.g.,~tables or visualizations) as shown in \mbox{Figure~\ref{fig:dashboard} (left).} All the visualizations are represented internally as data tables, as visualized in Figure~\ref{fig:dashboard}~(right), with the chart type recorded as part of the metadata. Each sub-panel is associated with a variety of captions describing the key insights; Table~\ref{tab:insights} shows the number of insights generated for each insight type and each sub-panel (\dashYel{A}~\dashOra{B}~\dashPin{C}~\dashTea{D}~\dashBlu{E}) for our sample dashboard (Figure~\ref{fig:dashboard}). For each individual insight, we record relevant properties (\eg the referenced data values, attributes, metrics, insight type, \etc) and pair this information with metadata from the dashboard (\eg the sub-panel position in the dashboard layout and information about the visualization type). We use this information to (1)~compute compound scores for the insights and (2)~generate links between insights in the network.

\bpstart{Insight Types} 
For demonstration purposes, we generate twelve types of insights leveraging an approach based on Voder~\cite{srinivasan-voder}: \emph{minimum}~(\textsc{mi}), \emph{maximum}~(\textsc{mx}), \emph{max extent}~(\textsc{me}), \emph{highest bar}~(\textsc{hb}), \emph{skew}~(\textsc{sk}), \emph{long tail distribution}~(\textsc{lt}), \emph{seasonality}~(\textsc{se}), \emph{trend}~(\textsc{tr}), \emph{spike}~(\textsc{sp}), \emph{decline}~(\textsc{de}), \emph{anomaly}~(\textsc{an}), and \emph{correlation}~(\textsc{co}).  We use these abbreviations throughout the paper and within our sample applications. Table~\ref{tab:insights} shows an example insight of each type for reference. However, it is important to note that our framework is not limited to these types of insights; instead, they were chosen for illustrative purposes alongside our sample dashboard (Figure~\ref{fig:dashboard}). Adding, removing, or changing the insight types would simply impact the number of nodes and links produced with our~framework.

Each insight is assigned a short ID: the first character denotes the visualization type (\textbf{\underline{B}}ar, \textbf{\underline{D}}onut, \textbf{\underline{L}}ine, \textbf{\underline{M}}ulti-line, or \textbf{\underline{T}}able); the next two characters are based on the topics (\ie metrics and dimensions) that are visualized; the fourth character refers to the column in the data table, or uses a dash (-) to denote the entire table; and the last two characters are the insight abbreviation as described above.
\subsection{Type-based Links}
\label{sec:type}

The first category of links includes general properties related to the underlying insight type, which defines \emph{how} or \emph{why} the insight was generated. For our sample dashboard (Figure~\ref{fig:dashboard}), we introduce three pieces of type-based information: (1)~\emph{insight type}, (2)~\emph{comparison type}, and (3)~\emph{chart~type}. These type-based links may capture details of the particular insight-generation approach and/or insight templates, and could thus vary for different instantiations of the insight network framework; for example, while it was not captured in our example dashboard scenario, a common extension would be to introduce links associated with the intended \emph{analysis task}.

The most direct form of type-based links encode the particular (1)~\emph{insight type}, as outlined in Section~\ref{sec:overview}; for example, a user may want to focus on analyzing all of the \emph{spikes} that occur in the data, regardless of which topic or sub-panel they come from, \eg \insightSPa\ and \insightSPb. Figure~\ref{fig:connections}\insightT\ shows insights clustered only by the insight type. It is important to note that the insight types depend on the generation approach; for example, we chose to differentiate between \emph{highest bar} (\textsc{hb}) and \emph{max} (\textsc{mx}) insights depending on the \emph{chart type}. As noted in Section~\ref{sec:overview}, our framework could also apply to different insight types or classifications~\cite{chen2009toward}, which would impact the overall number of links.

Perhaps similar to the classification of \emph{analysis task} or \emph{insight type}, the insights we generate often include a (2)~\emph{comparison type} (\eg total, more than, \etc) that describes the relationship that is encoded in the underlying values. For example, the insight \insightSPd compares the identified domain values (``Positive'' and ``Very Positive'') to the \emph{total} value for \texttt{Sentiment} (\ie \insight{22\% in total}).

As noted above, the (3)~\emph{chart type} may impact which insights are generated, and hence describes another relationship between the insights, albeit one which is less granular and thus leads to larger clusters. In Section~\ref{sec:overview}, we identified five different chart types as part of the unique ID (bar, donut, line, multi-line, and table); however, for our small example dashboard, we do not include repeating chart types, which results in separate categories for each sub-panel, similar to the layout-based links (see Section~\ref{sec:layout}). On the other hand, for our insight-generation approach, we leverage three categorizations of chart type to determine which insights are generated (bar \& donut, line, and multi-line); we then treat the tables as either bar charts or line charts depending on the data. We thus opt to use these chart categorizations as the chart type in our running example. Encoding other chart types simply impacts the number of~links.
\subsection{Topic-based Links}
\label{sec:topic}

The second link category describes the topics (\ie data attributes) that are represented in the insights. We identified two straightforward types of topic-based information in our generated insights: the underlying data (1)~\emph{metrics} or (2)~\emph{dimensions}. While there may be some overlap with the metadata-based links related to the dashboard layout (Section~\ref{sec:layout}), topic-based links directly co-locate insights that occur at different positions throughout the dashboard.

An important characteristic of topic-based links is that the underlying data (1)~\emph{metrics} or (2)~\emph{dimensions} may appear in multiple sub-panels of a dashboard, suggesting that the topic is of particular interest to the dashboard-creator. In our running example, \metricFirst\ is used in \metricFirstCount\ sub-panels (Figure~\ref{fig:dashboard}~\metricFirstPanels), whereas the \metricSecond\ is only used in \metricSecondCount\ sub-panels (Figure~\ref{fig:dashboard}~\metricSecondPanels); Figure~\ref{fig:connections} shows how the insights cluster for the metrics (Figure~\ref{fig:connections}\topicM) and dimensions (Figure~\ref{fig:connections}\topicD); notably, this clustering introduces duplicate nodes when an insight has multiple dimensions, such as \insight{`Positive' had the highest value, with 48.837 in `Avg. Duration' of `Call Reason' for  `Payments' (25\% in total)} (from sub-panel~\dashTea{D}).

Furthermore, topic-based links may also consider other features of the data such as filters or segments. \exampleTableFilters\ The resulting links are highly similar to the value-based links described in the next section; the key difference between topic-based and value-based links is whether the characteristic is shared by all of the generated insights for that sub-panel, or only a subset.
\subsection{Value-based Links}
\label{sec:value}
The third category expands on the idea of the topic-based links to provide more granular information about the particular values that are referenced in the insight; whereas the topic-based links generally apply to all of the insights associated with the same underlying data (independent of the exact data itself), the value-based links are representative of the particular trends that arise within the data.

One common example of this category includes dates and times, which can also provide an intuitive ordering to the insights. While singular date/time references are straightforward to cluster (and order), a notable complexity arises when working with both singular and ranged values. Consider the following two insights from sub-panel~\exampleTimePanel\,: \exampleTimeA\ and \exampleTimeB\ These insights refer to a related time-period, but different orderings may impact the attention the reader pays to each one. Figure~\ref{fig:connections} shows the insights approximately ordered by date; insights without a date are excluded from this sub-figure.

Another interesting value-based example is the comparison \emph{percentage}; as discussed in Section~\ref{sec:type}, many of the insights we generate include a comparison type (\eg total, more than, \etc); hence, all but one of our insights (\dashBlu{MCS-CO}) include a percentage value of some sort (see~Table~\ref{tab:insights}). We thus extract these percentage values to support a different type of connection; for example, the following insights both describe values accounting for \myquote{34\% in total,} while otherwise having a variety of different characteristics: \myquote{The values of `Calls' are highly skewed towards `Tuesday' and `Wednesday' (34\% in total)} (\dashOra{BCW-SK}) and \myquote{Negative had the greatest value, with 11,063 in `Calls' (34\% in total)} (\dashPin{DCS-MX} and \dashTea{TDX0MX}).
\subsection{Metadata-based Links}
\label{sec:layout}

The fourth category of links introduces relevant metadata-based information into the insight relationships. Chen~\etal describe some examples, such as the creation date or author~\cite{chen2009toward}; we expand upon this information and particularly emphasize metadata for our example scenario related to the underlying dashboard \emph{layout}.

The dashboard layout can capture some of the original intent of~the dashboard-creator in terms of prioritization, \eg sub-panels near the top may indicate that the information is more important or requires more frequent access; in some cases, the top charts act as an overview, followed by a more in-depth exploration of the data. Similarly, the layout in the underlying data table (see~Figure~\ref{fig:dashboard},~right) may provide information about the priority and/or evolution of calculated metrics, often in reading-order; for example, in Figure~\ref{fig:dashboard}~\dashTea{D}, the average call \texttt{Duration} ``Overall'' is shown \emph{after} the average call \texttt{Duration} for each call \texttt{Reason}.

We thus encode four distinct types of layout-related information within the category of metadata-based links. First, we include information about the position of the sub-panel within the dashboard based on both (1)~the \emph{panel row} and (2)~\emph{panel column}. For both the row and column (separately), we add a link between each pair of insights if the index matches. Figure~\ref{fig:connections}\layoutPR\ and Figure~\ref{fig:connections}\layoutPC\ illustrate the clusters (\ie cliques) produced by the \emph{panel row} and \emph{panel column} links, respectively. By combining these two link types in our cluster visualization, we can create a clustering of the insight nodes that mirrors the general layout of the dashboard (Figure~\ref{fig:connections}\layoutPRC).
When only the \emph{panel row} and \emph{panel column} links are active, insights will have an aggregated link weight of two if they were generated for exactly the same sub-panel in the underlying dashboard.

Next, we include links corresponding to the (3)~\emph{table column} in the underlying data table (Figure~\ref{fig:dashboard}, right). While this relationships is perhaps a bit odd in-and-of itself (Figure~\ref{fig:connections}\layoutTC), it can be combined with the \emph{panel row} and \emph{panel column} to further recreate the original dashboard layout in the clusters (Figure~\ref{fig:connections}\layoutC).

Finally, an important characteristic of the visual dashboard layout is the (4)~\emph{sort} attribute, \eg the dimension or metric that determines the sort order for the data. The \emph{sort} links denote if the data for the pair of insights is sorted by the same attribute. Interestingly, the \emph{sort} attribute might not be a topic that is described by the insight; for example, if the table in Figure~\ref{fig:dashboard}~\dashTea{D} was sorted by \texttt{Calls}, the insights corresponding to the average \texttt{Duration} for the ``Payments'' \texttt{Call Reason} would be sorted by an attribute that is \emph{not} described by the insight, and hence the insight data is \emph{not} sorted. These more complex relationships can be captured by combining this \emph{layout-based} information with \emph{topic-based} links (Section~\ref{sec:topic}) or through the specification of \emph{score-based} links~(Section~\ref{sec:score}) to determine if the particular data is visually sorted in the dashboard.

While some of these basic relationships from the dashboard layout will result in links that are fairly naive (e.g., the \emph{table column} links in Figure~\ref{fig:connections}\layoutTC), they can provide useful flexibility for capturing more complex or nuanced relationships when combined or aggregated with other links (\eg Figure~\ref{fig:connections}\layoutC). Furthermore, depending on the characteristics of the source dashboard, the layout-based links described here could be extended to include other relationships as appropriate, such as the \emph{table row} or other \emph{grouping} characteristics (\eg for different panels or tabs in the dashboard), among others.
\subsection{Score-based Links}
\label{sec:score}

The previous link categories relate to the intrinsic properties of the insights, and may thus fail to support general-purpose explorations of the data; for example, a user might wonder \textit{``what five insights are the most important overall?''} To support this form of exploration, we include a category of score-based or compound links. As a starting point for the compound links, consider Figure~\ref{fig:connections}\layoutPRC\ and Figure~\ref{fig:connections}\layoutC, which both use a combination of the \emph{layout-based} connections for the clustering. We can further extend this approach to create a score for each insight that combines the priority from the layout (\eg position, sorting, \etc) with a measure of how representative the insights are (\eg the prevalence of the values mentioned in the insight compared to all insights in the dashboard). 

While there are many ways such a score may be computed, consider the following example. We define a weighted ``priority'' score as: $priority = 0.3*layoutScore + 0.7*valueScore$, which gives a higher weight to the mentioned values compared to the overall layout. We can then define the $layoutScore$ as $0.25*panelRow + 0.25*panelCol + 0.5*tableCol$. The $panelRow$, $panelCol$, and $tableCol$ scores are all computed as the normalized reverse index value in the layout (\ie one is the first row/column and zero is the last row/column). Next, we can compute the $valueScore$ as follows: (1)~count the occurrences for all dimension \emph{values} across all of the insights, such that $c_x$ is the count of the value $x$; (2)~for each insight, compute the average occurrence as $vScore = \frac{1}{n} * \sum c_x$ where $n$ is the number of unique values mentioned in the insight; (3)~compute the $valueScore$ by adjusting the $vScore$ for the insight based on the $min$ or $max$ for all insights: $valueScore = (vScore - min(vScore)) / (max(vScore) - min(vScore))$. Note that the \emph{score-based} links can be extended to cover any unique formula or combination of other scores; for our example scenario, we include \numberScores\ scores: \texttt{priority}, \texttt{layoutScore}, and \texttt{valueScore}.

All of the links described in the insight network framework so far encode an exact match in the characteristics of the insights (\eg~denoting the same \emph{panel row}, \emph{dimensions}, \emph{dates}, \emph{values}, \emph{scores}, \etc); however, some of these characteristics may support additional ordered browsing \emph{between} categories (\eg moving from the first \emph{panel row} to the second \emph{panel row} is a stronger relationship than moving from the first to the fourth, then back to the second). Hence, we can leverage the score-based links (Section~\ref{sec:score}) to optionally add any of these expanded or weighted relationships to further configure the connection and exploration process as needed.

\newcommand{\sysNetwork}{A}
\newcommand{\sysStory}{B}
\newcommand{\sysSumPar}{c}
\newcommand{\sysNarEle}{d}
\newcommand{\sysHide}{e}

\begin{figure}[t]
    \centering
    \includegraphics[width=\columnwidth]{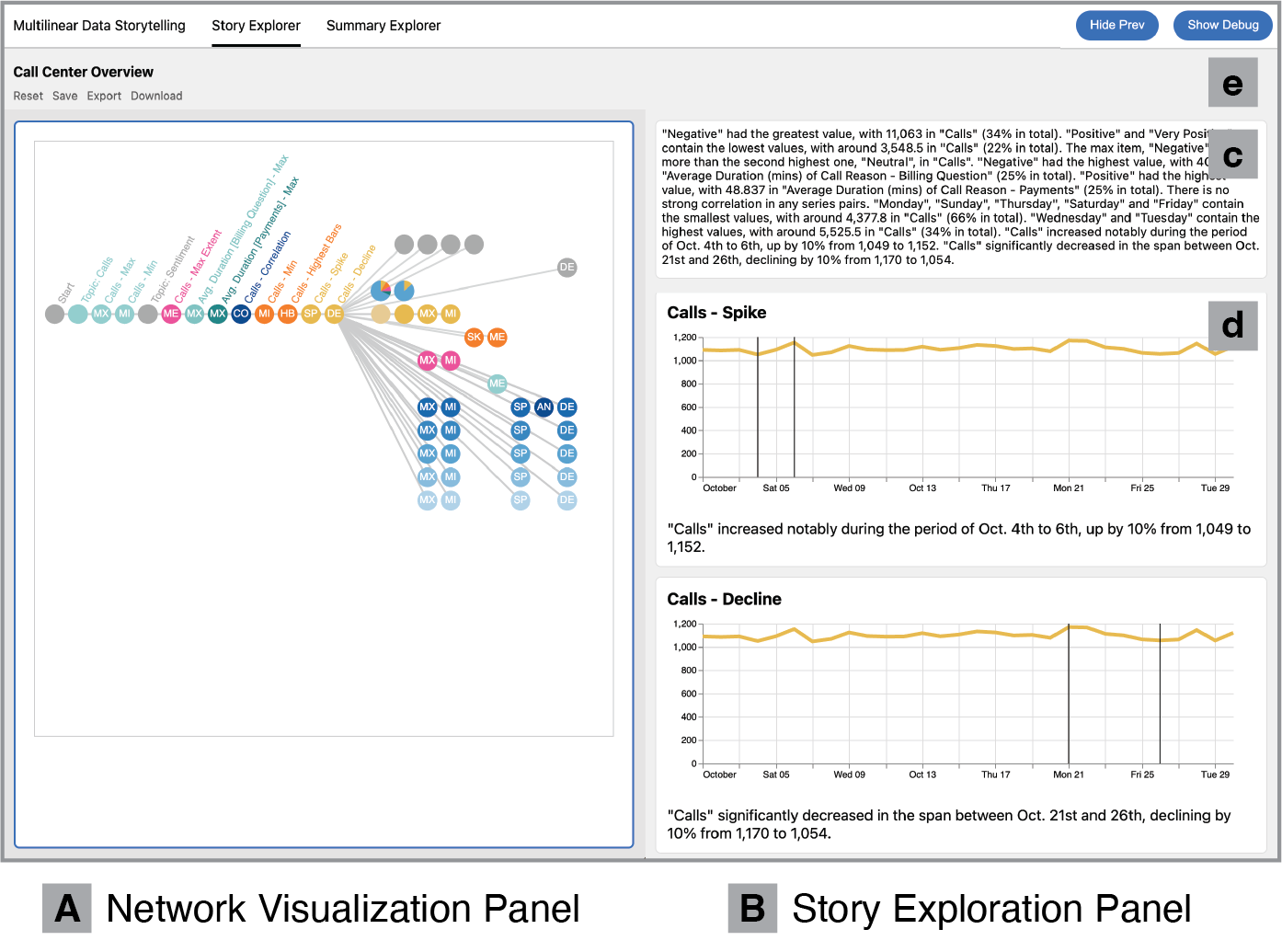}
    \caption{Our visualization prototype has two components: (\sysNetwork)~the interactive \textbf{visualization panel} supports exploration of linked insights using different visual representations (such as the node-link network view shown here); (\sysStory)~the \textbf{story exploration panel} includes (\sysSumPar)~a concatenated insight paragraph and (\sysNarEle)~linear narrative components containing the insights and accompanying visualizations.}
    \label{fig:system}
\end{figure}
\section{Visualizing the Insight Network}
\label{sec:system}

We developed a visualization playground to experiment with visual network representations to help conceptualize the relationships and scope of our insight~network framework. Another goal of this visualization playground is to demonstrate how interactive capabilities can support aggregation or simplification of the network, to inform the design of future applications; we explore one such LLM-based application as a case study in Section~\ref{sec:llm}. This playground is not intended as a standalone application nor as the ideal visual representation for end-users; instead, we have included versions of these visualizations throughout the paper for explanatory purposes. In this section, we reflect on the utility of the different visualizations for representing the complexities of our insight network framework. 

Our visualization playground has two interactive components: (1)~a \emph{visualization panel} (Figure~\ref{fig:system}\sysNetwork) that enables the user to interactively explore visual representations of the insight network and select insights and (2)~a \emph{linear story panel} (Figure~\ref{fig:system}\sysStory) that shows the concatenated insight paragraph for the selected insights and the individual story components (with the corresponding visualization). Our playground includes four visual representations: (1)~the \emph{dashboard}, (2)~a~\emph{network visualization} with a custom graph layout, (3) a node \emph{cluster visualization}, and (4) a link \emph{matrix visualization}.

%%%%%%%%%%%%%%%%%%%%%%%%%%%%%%%%%%%%%%%%%%%%%%%%%%%%%%%%%%%%%%%%%%%%%%%%%%%%%
\subsection{Network Visualization Panel}
\label{sec:network_vis}

The network visualization acts as a direct node-link representation of our insight network framework to enable user-guided exploration of the underlying insights. To facilitate exploration and simplify the network, we introduce aggregated ``gatekeeping'' nodes to represent the different categories of connections encoded in our insight network framework. We also leverage a custom graph layout to organize insights similar to the cluster visualization (Section~\ref{sec:cluster}); we can toggle between the force-directed and custom layout. For more detail on the implementation of the aggregate gatekeeping nodes and custom layout, please see the supplemental material.

For illustrative purposes, the nodes are colored based on the corresponding sub-panel from the source dashboard; different shades are used to indicate different dimensions or metrics in the sub-panel. We assign the colors based on the \texttt{layoutScore} described in Section~\ref{sec:score}; the color is also configurable in our playground. Insight nodes include a white text label abbreviating the insight type, as described in Section~\ref{sec:overview}. Finally, we optionally support aggregation of the links similar to the matrix visualization (Section~\ref{sec:matrix}), with the stroke weight encoding the number of links.

%%%%%%%%%%%%%%%%%%%%%%%%%%%%%%%%%%%%%%%%%%%%%%%%%%%%%%%%%%%%%%%%%%%%%%%%%%%%%
\subsection{Node Cluster Visualization Panel}
\label{sec:cluster}

The cluster visualization focuses on grouping related insights based on one or more characteristics from our insight network framework, as shown in Figure~\ref{fig:connections}. These clusters represent cliques wherein all nodes are connected to all other nodes in the cluster for the chosen link characteristic(s). It is important to note that this representation \emph{does not} show all possible links encoded in the underlying insight network framework, only the chosen relationships. This view also produces \emph{duplicate} nodes if the insight corresponds to multiple clusters; we highlight these duplicate nodes on mouseover.

Our cluster visualization supports up to four link categories~(two for each axis); additional customization of this visualization can be achieved by creating custom score-based links in the underlying insight network framework. Most of the cluster views shown in Figure~\ref{fig:connections} show only one link type at a time; Figure~\ref{fig:connections}\layoutPRC\ visualizes the \emph{panel row} as the row and \emph{panel column} as the column, thereby filtering to two link types; similarly, Figure~\ref{fig:connections}\layoutC\ goes a step further to encode cliques sharing all three link types (\emph{panel row}, \emph{panel column},~and \emph{table column}). The cluster visualization includes axis labels for the row and column to illustrate the encoded relationships.

\begin{figure}[t]
    \centering
    \includegraphics[width=0.9\columnwidth]{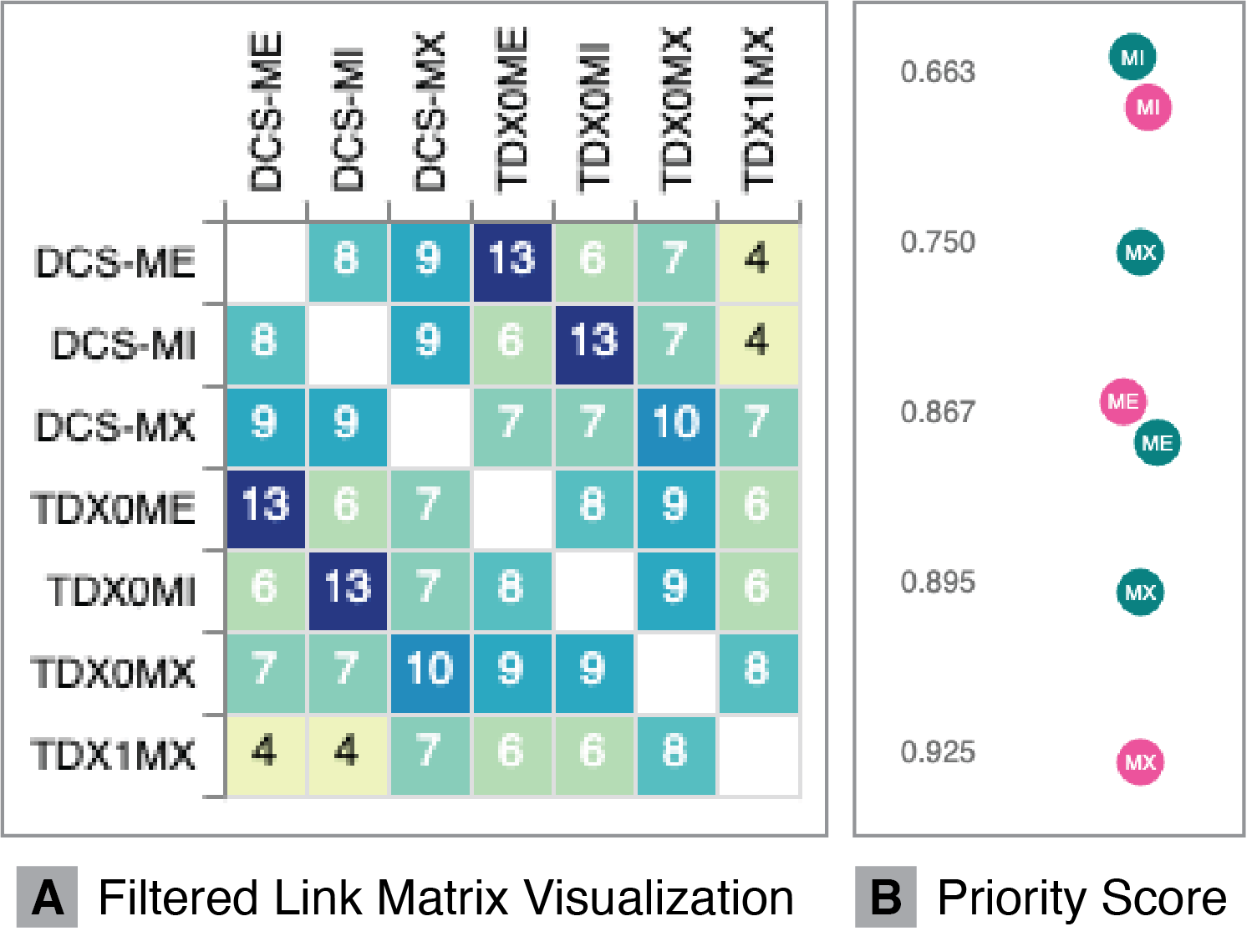}
    \vspace{-5px}
    \caption{(A) The matrix visualization and (B) cluster visualization of the \texttt{priority} score for the subest of seven insights in Figure~\ref{fig:teaser}.}
    \label{fig:matrix}
    \vspace{-10px}
\end{figure}

%%%%%%%%%%%%%%%%%%%%%%%%%%%%%%%%%%%%%%%%%%%%%%%%%%%%%%%%%%%%%%%%%%%%%%%%%%%%%
\subsection{Link Matrix Visualization Panel}
\label{sec:matrix}

The matrix visualization emphasizes the total number of links between each pair of insights in our insight network framework. The insight ID is displayed along the x and y axes to create a heatmap encoding the aggregate number of links. Figure~\ref{fig:matrix}A shows the matrix visualization for the subset of seven insights selected in Figure~\ref{fig:teaser}); insight \dashPin{DCS-MX} has the highest priority score, and the following insights are sorted first by the number of links in the matrix (Figure~\ref{fig:matrix}A), and second by the priority score (Figure~\ref{fig:matrix}B).  The full matrix visualization for our example dashboard (Figure~\ref{fig:dashboard}) is included in the supplemental material. To facilitate browsing, our visualization playground supports interactive filtering of the link categories to highlight the different connection patterns that arise.

%%%%%%%%%%%%%%%%%%%%%%%%%%%%%%%%%%%%%%%%%%%%%%%%%%%%%%%%%%%%%%%%%%%%%%%%%%%%%
\subsection{Linear Story Exploration Panel}

The story panel has two key components: (1)~a single paragraph displaying the selected insights (Figure~\ref{fig:system}\sysSumPar) and (2)~an expanded narrative containing the individual story components (Figure~\ref{fig:system}\sysNarEle). When the user selects a new insight, the raw insight text is appended to the overview paragraph and an annotated story component (containing the title, insight text, and representative visualization) is displayed in the expanded narrative. To create the visualizations, we leverage a set of custom Vega-Lite~\cite{satyanarayan-vegalite} templates based on the chart type originally specified for each sub-panel in the dashboard; the visualization also highlights any values mentioned in the insight.

For example, when the user selects the insight node corresponding to \selectedNode\ in Figure~\ref{fig:system}\sysNetwork, the insight is appended as the last sentence of the summary (Figure~\ref{fig:system}\sysSumPar), \selectedInsight\ and a new story component is added to the expanded narrative which shows a \selectedVis\ with the values \selectedVisHighlight highlighted (Figure~\ref{fig:system}\sysNarEle).

%%%%%%%%%%%%%%%%%%%%%%%%%%%%%%%%%%%%%%%%%%%%%%%%%%%%%%%%%%%%%%%%%%%%%%%%%%%%%
\subsection{Reflections on Visualizing the Insight Network}
\label{sec:network-reflections}

The goal of this visualization playground was to illustrate the characteristics and configurability of our insight network framework. 
While the visualization playground is not intended to be an ideal, standalone application, such visualizations and/or interaction techniques could be integrated with analytics tools alongside the source dashboard to provide a more direct connection between the different components. Our insight network framework provides a structured way to navigate the relationships between individual insights, and can thus support the design of different visualizations for different use cases, as demonstrated here. Furthermore, our framework does not fundamentally require a visual representation; instead, the relationships encoded can be used as the underlying data representation for other applications, as shown in the next section. 
\section{Case Study: LLM-based Summary Generation}
\label{sec:llm}

To illustrate the utility of our dense insight network framework,~we developed an LLM-based summary generation application as a case study; this development process was done in collaboration with several expert stakeholders, as described in Section~\ref{sec:participatory}. Based on iterative feedback from our expert stakeholders, we found that the process of writing dashboard summaries often requires enumerating all possible insights in order to select the ones that are most important. Our stakeholders then refine the chosen insights to produce a summary paragraph to share with a broader audience.

We implement our case study application in Python; we first encode the dashboard and insights using our insight network framework, and refine the \texttt{priority} score (Section~\ref{sec:score}) based on feedback from our expert stakeholders. Using these \emph{score-based} links, we select the top insights, order them into paragraphs for each sub-panel (\ie based on the layout properties encoded in the \emph{metadata-based} links), and use an LLM to generate a more concise summary.

This LLM-based summary generation process was designed to address two key concerns: (1)~by ranking and ordering the selected insights from the much larger collection of all possible insights,~we aim to reduce the risk of the LLM missing the most important information while also providing more explainability and control around the selection process; and (2)~through several rounds of iteration, we perform prompt engineering to settle on an approach that generates concise, natural-sounding summaries and share our findings.

\subsection{Iterative Design Feedback}
\label{sec:participatory}

To develop our case study application, we elicited feedback from a set of expert stakeholders. For approximately five months, we~had weekly meetings with between one to three experts who develop and analyze dashboards and analytics tools everyday. These meetings aimed to understand experts' requirements and thought process on ranking and summarizing automatically-generated insights.

During these weekly meetings, we iteratively shared sets of generated insights and the selection procedures used according to our insight network framework. When available, we leveraged example dashboards provided by our expert stakeholders to produce concrete results showcasing the role of different network characteristics on the selection and summarization procedure. Leveraging the insight network framework allowed us to quickly reconfigure and extract different insight combinations for discussion throughout the development process. Based on our experts' feedback, we iteratively refined the features of our insight network to develop a custom, compound \emph{priority} score (Section~\ref{sec:score}) for our application.

Throughout the development process, we welcomed any and all feedback from our expert stakeholders on which properties to encode and explore for the final summarization methodology. While some of the characteristics (\ie links) we discussed were not ultimately incorporated in the final summarization approach for this sample application, the conversations were still useful for demonstrating the utility of our insight network framework for encoding and exploring complex relationships between dashboard insights.

%%%%%%%%%%%%%%%%%%%%%%%%%%%%%%%%%%%%%%%%%%%%%%%%%%%%%%%%%%%%%%%%%%%%%%%%%%%%%
\subsection{Ranking and Ordering Selected Insights}
\label{sec:ordering}

Leveraging our insight network framework, we iteratively select the top scoring insight based on the \texttt{priority} score described in Section~\ref{sec:score}. We also select all other insights with the same score, and continue the selection process with the next highest scoring insight until we have the desired number of insights for the summary. For this paper, we aim for between four and fifteen insights as the target number for summarization. We then reorder the insights from the \emph{score-based} order into a \emph{metadata-based} order such that insights are grouped approximately by the layout and topic when sent as input to the LLM; an example of this ordering is included in the next section along with a discussion of our prompting strategy. 

%%%%%%%%%%%%%%%%%%%%%%%%%%%%%%%%%%%%%%%%%%%%%%%%%%%%%%%%%%%%%%%%%%%%%%%%%%%%%
\subsection{Summary Prompt Engineering}
\label{sec:prompt}

For this work, we use OpenAI's GPT-3.5~(\emph{gpt-35-turbo-v0613}) as our backbone LLM~\cite{NEURIPS2020_1457c0d6}. In order to strike a balance between readability and factuality, we choose a decoding temperature of 0.5 as a mid-point between the default temperature of 0.7, used for more creative generation, and the deterministic temperature of 0.0, commonly used for Q\&A~\cite{ho2023wikiwhy}. As for~the generation length, we allow for the same number of tokens as the total tokens in the selected insights, with the prompt itself expressing our preference for fewer sentences. Our final prompt is included below; to provide a more complete example, 
we use the six insights that were automatically selected for our example~dashboard (Figure~\ref{fig:dashboard}).\\\\

% \vspace{10px}
\begin{mdframed}
\footnotesize
%\scriptsize
\texttt{%
Write a summary of the data report below\\ using one third of the sentences.\\\\
Report:}\\
\footnotesize

\vspace{-4px}\noindent\textbf{About:} \emph{``Calls'' topped at 1,170 on Oct. 21st. It was 7\% more than the\,average\,of\,1,098.}\\

\vspace{-4px}\noindent\textbf{About:} \emph{``Wednesday'' and ``Tuesday'' contain the greatest values, with around 5,525.5 in ``Calls'' (34\% in total). The values of ``Calls'' are highly skewed towards ``Tuesday'' and ``Wednesday'' (34\% in total). The max item, ``Wednesday'', is 2\% more than the second highest one, ``Tuesday'', in ``Calls''.}\\

\vspace{-4px}\noindent\textbf{About:} \emph{``Negative'' had the greatest value, with 11,063 in ``Calls'' (34\% in total).}\\

\vspace{-4px}\noindent\textbf{About:} \emph{``Negative'' had the greatest value, with 11,063 in ``Calls'' (34\% in total).}\\\\
% \scriptsize
\texttt{Summary:}
\end{mdframed}

\vspace{10px}\noindent The resulting LLM-based summary is as follows: \myquote{On October 21st, the highest number of calls reached 1,170, which was 7\% higher than the average of 1,098. The days with the highest call volume were Wednesday and Tuesday, accounting for 34\% of the total calls. The values of calls were highly skewed towards these two days. Wednesday had the highest number of calls, which was 2\% higher than Tuesday. The highest value in the negative category for calls was 11,063, accounting for 34\% of the total calls.}

%%%%%%%%%%%%%%%%%%%%%%%%%%%%%%%%%%%%%%%%%%%%%%%%%%%%%%%%%%%%%%%%%%%%%%%%%%%%%
\subsection{Reflections on LLM-based Summary Generation}
\label{sec:network-reflections}

Compared to direct concatenation of the selected insights, there are several advantages to employing an LLM-based prompting strategy: first, the paraphrased insights can provide more variety in the sentence structure or more natural-sounding summaries compared to the original, template-generated insights; second, the LLM may combine or reduce repetitive information from across multiple insights. For example, several identical insights are generated for the donut chart~\dashPin{C} and table~\dashTea{D} in our example \mbox{dashboard (Figure~\ref{fig:dashboard})} due to the same data appearing in multiple locations, hence the repeated insights in our sample prompt in Section~\ref{sec:prompt}. While simple deduplication could remove these direct redundancies, the LLM-based summarization can adjust the output to change an input insight like \myquote{The values of `Calls' are highly skewed towards `Tuesday' and `Wednesday'\,\hspace{-2px}} (\dashOra{BCW-SK}) to a rephrased sentence like \myquote{The values of calls were highly skewed towards these two days,} which takes advantage of the earlier sentences in the summary paragraph to provide the context (\ie \myquote{The days with the highest call volume were Wednesday and Tuesday}). However, a disadvantage of such an approach is the possibility of hallucination; we provide a deeper discussion of such concerns in Section~\ref{sec:hallucination}.
\section{Limitations and Future Work}

We discuss several limitations, as well as future work on the evaluation of new insight selection and summarization approaches.

\subsection{Hallucinations in LLM-based Summaries}
\label{sec:hallucination}

A major concern for our case study application (Section~\ref{sec:llm}), as well as LLM-based approaches more generally, is the potential for hallucination in the results. To combat this concern, we use the insight network framework for the initial insight selection process, which reduces our reliance on the behavior of the LLM when it comes to identifying the most important insights. By leveraging the insight network framework, we can better control the selection process and produce more explainable results to highlight how and why insights were selected. However, hallucination is still a concern, even for the limited role it plays in our approach. As discussed in Section~\ref{sec:llm}, we performed several rounds of iteration on the prompt; one version included the chart title along with each paragraph, as shown below:

\vspace{10px}
\begin{mdframed}
\footnotesize
\texttt{%
Write a summary of the data report below using one third of the sentences.\\\\
Report:}\\
\footnotesize

\vspace{-4px}\noindent\textbf{About Number of Calls Over Time:} \emph{``Calls'' topped at 1,170 on Oct. 21st. It was 7\% more than the average of 1,098.}\\

\vspace{-4px}\noindent\textbf{About Number of Calls per Day of the Week:} \emph{``Wednesday''~and ``Tuesday'' contain the greatest values, with around 5,525.5 in ``Calls'' (34\% in total). The values of ``Calls'' are highly skewed towards ``Tuesday'' and ``Wednesday'' (34\% in total). The max item, ``Wednesday'', is 2\% more than the second highest one, ``Tuesday'', in ``Calls''.}\\

\vspace{-4px}\noindent\textbf{About Total Calls by Sentiment:} \emph{``Negative'' had the greatest value, with 11,063 in ``Calls'' (34\% in total).}\\

\vspace{-4px}\noindent\textbf{About Average Duration by Call Reason and Sentiment:} \emph{``Negative'' had the greatest value, with 11,063 in ``Calls'' (34\% in total).}\\\\
\texttt{Summary:}
\end{mdframed}

\vspace{10px}\noindent The resulting LLM-based summary includes a subtle, but important hallucination; keep an eye out for it as you read through the following paragraph: \myquote{The highest number of calls, reaching 1,170, was recorded on Oct. 21st, which was 7\% higher than the average of 1,098. Wednesday and Tuesday had the highest number of calls, totaling around 5,525.5, accounting for 34\% of the total calls. The distribution of calls is highly skewed towards Tuesday and Wednesday. The highest sentiment category was ``Negative'' with 11,063 calls, accounting for 34\% of the total. Similarly, ``Negative'' had the highest average duration for call reasons and sentiments.}

Notably, the final sentence, which states that
\myquote{`Negative' had the highest average duration for call reasons and sentiments,} replaces the true insight (\dashTea{TDX0MX}\,, \ie \myquote{the greatest value... in `Calls'\,}) with a mention of the \myquote{Average Duration by Call Reason and Sentiment} from the title of the table in Figure~\ref{fig:dashboard}\,\dashTea{D}. This error is somewhat difficult to spot when reading the summary paragraph, because the data in the original dashboard \textit{almost} exhibits the same information (\eg ``Negative'' does in fact have the highest average duration overall, as well as the highest for all call reasons other than ``Payments''). However, the LLM was not given this knowledge, and hence hallucinates only based on the chart title, which results in a fairly confusing sentence when performing a close reading.

\subsection{Next Steps: Large Scale Summary Evaluation}
\label{sec:future-evaluation}

Motivated by the hallucination example from the previous section, along with other observations from our case study application, in future work we plan to leverage our insight network framework as the foundation for evaluating different natural language summaries; an important part of such evaluations will be to understand and explore the nuance of hallucinations for various prompting strategies.

We would also like to explore additional evaluation metrics beyond the presence or impact of LLM-based hallucinations.
In particular, we identified two research questions related to insight selection and summarization: (1)~\emph{how consistent are people for identifying the most important insights?} Our case study application uses the insight network framework to provide control and explainability around what insights are selected, rather than relying solely on an LLM; however, different people may have different strategies or priorities for selecting insights, which we would like to further explore. We plan to leverage our insight network framework to provide a stronger foundation for analyzing the key characteristics that arise in this selection process. For our second research question, we would also like to better understand (2)~\emph{how do human judges assess the quality of dashboard summaries?} We believe that leveraging the insight network framework can provide an interesting foundation to assess the focus and intent of both auto-generated and user-authored summaries. As part of this evaluation, we also hope to better understand what other characteristics of the insights or resulting summaries are most relevant to the evaluation process.

\section{Conclusion}
In this paper, we contributed an insight network framework that includes five high-level categories of links between automatically-generated dashboard insights. We developed a visualization playground to illustrate both the complexity and flexibility of our framework, and to further demonstrate potential interactive capabilities for simplifying the representation; we then showed how this framework can support the development of future applications via a case study for an LLM-based summarization application that supports generating, scoring, ranking, and selecting insights to summarize complex dashboards using concise, natural language.

%-------------------------------------------------------------------------
% bibtex
\bibliographystyle{eg-alpha-doi} 
\bibliography{sections/bibliography}       

% biblatex with biber
% \printbibliography

\end{document}